\documentclass{PoS}
\usepackage{wrapfig}

\title{Clustering in nuclei from {\em ab initio} nuclear lattice simulations}

\ShortTitle{Clustering in nuclei from ab initio nuclear lattice simulations}

\author{\speaker{Ulf-G. Mei{\ss}ner}%
        \thanks{Work supported in part by DFG and NSFC (CRC 110), the HGF Virtual Institute NAVI (VH-VI-417) and the Chinese Academy of 
        Sciences, PIFI grant 2015VMA076.}\\
       Helmholtz-Institut f\"ur Strahlen- und Kernphysik and Bethe Center for Theoretical Physics, Universit\"at Bonn, 53115 Bonn, Germany \\
       and\\ 
       Institute for Advanced Simulation, Institut f\"ur Kernphysik, J\"ulich Center for Hadron Physics, JARA-HPC and JARA-FAME, 
       Forschungszentrum  J\"ulich, 52425 J\"ulich, Germany\\
       E-mail: \email{meissner@hiskp.uni-bonn.de}}

\abstract{Nuclear Lattice Effective Field Theory is a new many-body approach that 
          is firmly rooted in the symmetries of QCD. In particular, it allows for 
          truly {\it ab initio} calculations of nuclear structure and reactions.
          In this talk, I focus on the emergence of $\alpha$-clustering in nuclei
          based on this
          approach. I also discuss various recent achievements, the deficiencies
          of the chiral forces used at present and the prospects to improve upon
          these and the calculations of nuclear properties and dynamics.}

\FullConference{The 8th International Workshop on Chiral Dynamics, CD2015\\
		29 June 2015 - 03 July 2015\\
		Pisa,Italy}

\begin{document}

\section{Short introduction}

Clustering is one of the most fascinating and intricate features of nuclear physics. A special role is played by the $\alpha$-particle,
which is not only very strongly bound ($E/A \simeq 7.1\,$MeV) but also spin- and isospin-saturated. While the theoretical foundations
for clustering in nuclei have already been laid out in the seminal work of Wheeler \cite{Wheeler:1937zza}, there is still lots of
on-going research on this phenomenon. Here, I just mention a few  recent works related to the $\alpha$-cluster structure in nuclei, namely the purely 
geometrical approach of Ref.~\cite{Bijker:2014tka} to explain the spectrum of $^{16}$O~\footnote{This work was indeed inspired by the
nuclear lattice simulation results reported below.}, the work on rod-like alpha structures in carbon isotopes at extreme spin and
isospin \cite{Zhao:2014vfa} or the work of Ref.~\cite{Ebran:2014pda} that performed systematic studies of
$\alpha$-clustering  based on density functional theory. Still, one important question remains unanswered: Can we understand this
phenonenom from {\em ab initio} studies? By this I mean a framework in which the two- and three-nucleon forces  are fixed in few-nucleon 
systems and based on these forces, the nuclear $A$-body problem is solved exactly and clustering emerges naturally. As I will show,
that it is indeed possible by combining the so successful nuclear chiral Effective Field Theory (EFT) (for a review, see \cite{Epelbaum:2008ga}, 
and the talk by Epelbaum for the most recent developments \cite{EE})
with lattice simulations, from here on called Nuclear Lattice EFT (NLEFT) or nuclear lattice simulations, for short.

\section{Basics of nuclear lattice simulations}

\begin{figure}[h]
\vspace{4mm}
\includegraphics[width=0.48\textwidth]{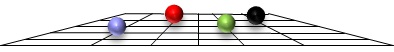}\hspace{1pc}
\includegraphics[width=0.48\textwidth]{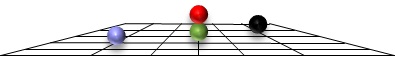}\\
\includegraphics[width=0.48\textwidth]{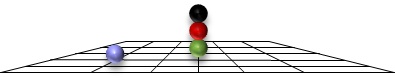}\hspace{1pc}
\includegraphics[width=0.48\textwidth]{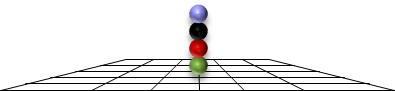}

\vspace{-2mm}
\caption{\label{fig:configs} Topologically different configurations
to put nucleons on a lattice. From the left upper to the lower right 
corner: all four nucleons on different sites, two/three/four nucleons
on one site, respectively.}
%
\end{figure}

In this section, I briefly review the main ingredients of nuclear lattice simulations, omitting all technical details.
The basic ingredient in this framework is the discretization of space-time, see Ref.~\cite{Borasoy:2006qn} for details. 
Space-time is represented by a grid.  This lattice serves  as a tool to compute the nuclear matrix elements under consideration. 
As a first step, we have to perform a Wick rotation  in time so that the time evolution operator behaves as $\exp(-Ht)$, with $H$ the 
nuclear Hamiltonian derived within chiral nuclear EFT, and $t$ is the Euclidean time. As a result,
space-time is  coarse-grained. In the  three spatial directions, the smallest distance  on the lattice is given by the lattice spacing $a$, so that 
the four-dimensional volume is $L\times L\times L\times L_t$,  with $L = N a$ and $N$ an integer, whereas in the time direction one uses a 
different spacing $a_t$, and  $L_t = N_t a_t$ is chosen as large as possible. Typical values are $N = 6...8$ and $N_t = 10...15$. 
As the Euclidean time becomes very large,  one filters out the ground state as it has the slowest fall-off with Eucldean time $\sim \exp(-E_0t)$, with $E_0$ the 
ground state energy. Excited states can also be 
investigated. This, however, requires some more effort, but it is considerably  easier than
extracting excited states in lattice QCD.  
The nucleons are placed on the lattice sites. Their interactions are given by one- and two-pion
exchanges and multi-nucleon operators, properly represented in lattice variables.  So far, nuclear lattice simulations have been carried out 
at next-to-next-to-leading order (NNLO) in the chiral expansion, incorporating two- and three-nucleon forces. The Coulomb force 
between protons and isospin-breaking  strong interaction effects are also included, thus one has all the
required ingredients to describe the structure of nuclei. The lattice is used to perform 
a numerically exact solution of the $A$-nucleon system, where the atomic number $A = N + Z$ counts the number of neutrons and protons in the nucleus 
under investigation. It is important to realize that the finite lattice spacing entails a maximum momentum $p_{\rm max} = \pi/a$, so that for a typical 
lattice spacing of $a = 2\,$fm, one has a maximum momentum of about 314~MeV. This means that  one deals with a very soft interaction. 
The main advantage  of this approach is, however, the approximate spin-isospin SU(4) symmetry of the nuclear interactions already realized by Wigner in 
1937~\cite{Wigner:1936dx}. Because of this approximate symmetry, the malicious sign oscillations that plague any fermion simulation at finite density
are very much suppressed,  quite in contrast to the situation in lattice QCD. A lucid discussion of this issue has been given in Ref.~ \cite{Chen:2004rq}.  Consequently, $\alpha$-type nuclei with $N = Z$, and spin and isospin zero can be simulated most easily. However, in the 
mean time our collaboration has developed a method that allows for a remarkable suppression of the remaining sign oscillations in nuclei 
with $N \neq Z$, as discussed below.  One more important  remark on the formalism is in order. The simulation algorithms sample all 
possible configurations of nucleons, in  particular one can have  up to four nucleons on one lattice site, see Fig.~\ref{fig:configs}. Thus, the so important phenomenon of clustering  in nuclei arises quite naturally  in this novel many-body approach, as discussed in more detail later. 

Most of the results presented in this talk have been obtained with the following lattice set-up: $a = 1.97\,$fm, $N = 7$, $a_t = 1.32\,$fm.
The forces have been obtained at NNLO, with 9 two-nucleon low-energy constants (LECs) fixed from fits to S- and P-wave $np$ phase shifts,
2 isospin-breaking NN LECs determined from the $nn$ and $pp$ scattering lengths and two 3N LECs fixed  from 
the triton binding energy and the axial-vector contribution to triton $\beta$-decay~\cite{Gazit:2008ma}.
Further, there is some smearing required in the LO S-wave four-nucleon terms with its size parameter determined from the
average $np$ S-wave effective range~\cite{Borasoy:2006qn}.

\section{Results from nuclear lattice simulations}

$\alpha$-clustering is known to play an important role in the carbon nucleus $^{12}$C, see e.g. Refs.~\cite{Tohsaki:2001an,Bijker:2002ac,Chernykh:2007zz,Freer:2014qoa}.
Ground-breaking studies using nuclear lattice simulations  have been performed for the spectrum and structure of $^{12}$C \cite{Epelbaum:2012qn},
presenting the first {\em ab initio} calculation of the so-called Hoyle state \cite{Epelbaum:2011md}, that plays a decisive  role in the  generation of carbon 
and oxygen in hot old stars. Further, in this framework it is possible to calculate the dependence of the various fine-tunings in the triple-$\alpha$
process as a function of the fundamental parameters of the Standard Model, in particular of the  light quark masses and the electromagnetic 
fine-structure constant \cite{Epelbaum:2012iu,Epelbaum:2013wla}. A detailed discussion of these issues and the relation to the anthropic
principle have been given in Ref.~\cite{Meissner:2014pma}. In the following, I concentrate on more recent results concerning the spectrum and
structure of $^{16}$O \cite{Epelbaum:2013paa} and the ground-state energies of the $\alpha$-type nuclei up to $^{28}$Si~\cite{Lahde:2013uqa}.
I further discuss a method \cite{Lahde:2015ona} that allows to tame the remaining sign oscillations, so that systematic calculations of halo nuclei 
or the exploration of the limits of nuclear stability become possible.

\begin{wrapfigure}{r}{0.5\textwidth}
  \begin{center}
\includegraphics[width=0.22\textwidth]{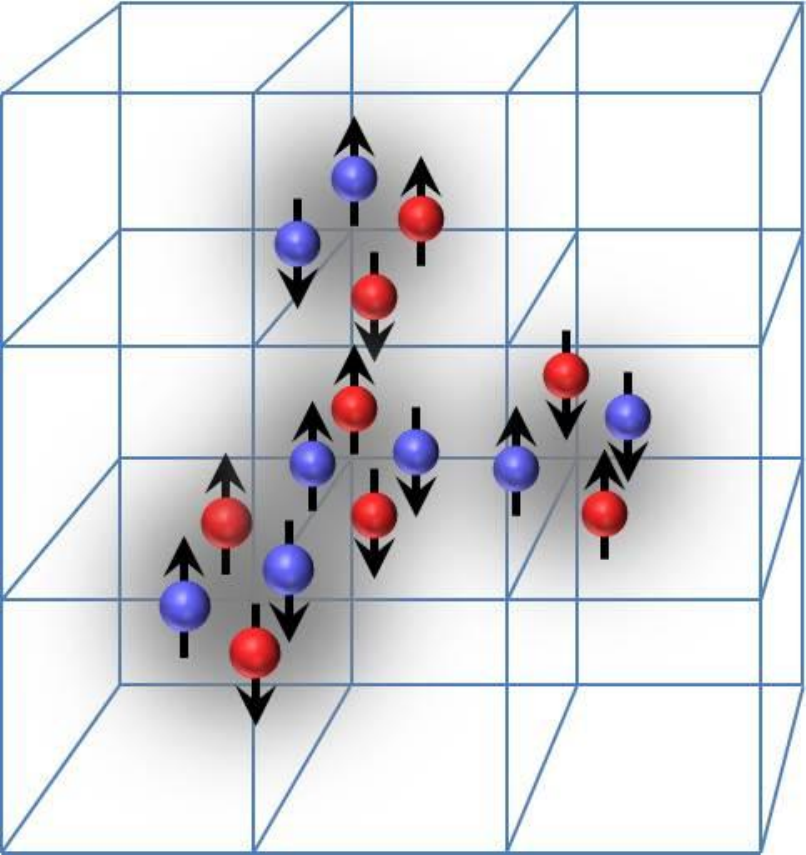}\hspace{1pc}
\includegraphics[width=0.22\textwidth]{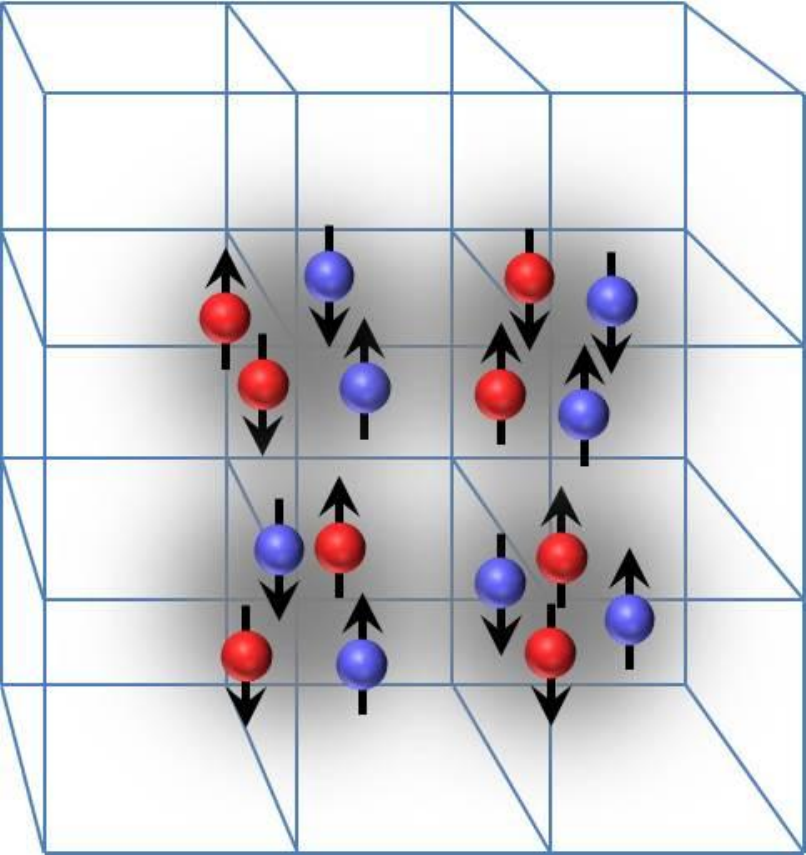}\\
\end{center}
\vspace{-2mm}
\caption{\label{fig:16O} 
Tetrahedral (left) and square (right) configurations in $^{16}$O.}
\vspace{-2mm}
\end{wrapfigure}

An important  $\alpha$ cluster-type nucleus is $^{16}$O, that also plays an important role in the formation of life on Earth.
Since the early work of Wheeler \cite{Wheeler:1937zza}, there have been several theoretical and experimental works that lend further 
credit to the cluster structure of $^{16}$O, see e.g. 
Refs.~\cite{Robson:1979zz,Bauhoff:1984zza,Tohsaki:2001an,Freer:2005ia}. 
However,  no {\em ab initio} calculation existed that gave further support to these ideas. 
This gap was filled in Ref.~\cite{Epelbaum:2013paa} where nuclear lattice simulations have been used to investigate the low-lying 
even-parity spectrum and the structure of the ground and first few excited states.  It is found that in the spin-0 ground state the nucleons are 
arranged in a tetrahedral configuration of $\alpha$-clusters, cf. Fig.~\ref{fig:16O}. For the first $0^+$ excited state, the predominant structure 
is given by a square arrangement of $\alpha$-clusters, as also shown in Fig.~\ref{fig:16O}. There are also rotational excitations of this square 
configurations that include the lowest $2^+$ excited state. Note, however, that these
snapshots of the wave functions depend on the lattice spacing $a$.

These cluster configurations can be obtained in two ways. First, one can investigate
 the time evolution of the various cluster configurations and  extract e.g. the corresponding energies as the Euclidean time goes to infinity. 
 Second, one can also 
 start with initial states that have no clustering at all. One can then measure the four-nucleon correlations. For such initial states, this density grows quickly
 with time and reaches a high level. For the cluster initial states, these correlations start out at a high level and stay large as a function of Euclidean 
 time. This is a clear indication that the observed clustering is not built in by hand but rather follows from the strong four-nucleon correlations in 
 the $^{16}$O nucleus. Or, stated differently, if one starts with an initial wave function
without  any clustering, on a short time scale clusters will form and make up the
 most important contributions to the structure of $^{16}$O or any type of nucleus, where $\alpha$-clustering is relevant. We have also performed investigation of the response of $^{16}$O 
to electromagnetic
probes. In Table~\ref{tab_trans} the NLEFT results at leading order (LO)
for the electric quadrupole moment of the $2^+_1$ state, the electric quadrupole ($E2$) 
transition probabilities, and the electric monopole ($E0$) matrix element are given.  Since the LO
charge radius $r_{\rm{LO}}^{}$ of the ground state is smaller than the 
empirical value $r_{\rm{exp}}^{}$, a systematic deviation appears, which arises from the overall size of the second moment of the charge distribution. 
To compensate for this overall scaling mismatch,  ``rescaled'' quantities multiplied by powers of the ratio 
$r_{\rm{exp}}^{}/r_{\rm{LO}}^{}$, according to the length dimension of each observable, were also calculated.
\begin{table}[h]
\begin{center}
\begin{tabular}{| c | c c | c |}
\hline  
& LO & rescaled & Exp. \\ \hline\hline
$r(0^+_1)$ [fm] & 2.3(1) & --- & 2.710(15) \cite{Kim:1978st} \\
$Q(2^+_1)$ [$e$ fm$^2$] & 10(2) & 15(3) & --- \\
$B(E2, 2^+_1 \to 0^+_2)$ [$e^2$fm$^4$] & 22(4) & 46(8) & 65(7) \cite{Ajzenberg:1971}\\
$B(E2, 2^+_1 \to 0^+_1)$ [$e^2$fm$^4$] & 3.0(7) & 6.2(1.6) & 7.4(2) \cite{Moreh:1985zz}\\
$M(E0, 0^+_2 \to 0^+_1)$ [$e$ fm$^2$] & 2.1(7) & 3.0(1.4) & 3.6(2) \cite{Miska:1975} \\
\hline 
\end{tabular}
\caption{NLEFT results for the charge radius $r$ of the ground state of $^{16}$O, 
the quadrupole moment $Q$, and the electromagnetic transition amplitudes for $E2$ and 
$E0$ transitions, as defined in Ref.~\cite{Mottelson_book}.
We compare with empirical (Exp.) values where these are known. For the quadrupole moment and the transition amplitudes, we also show
``rescaled'' LO results, which correct for the deviation from the empirical value of the charge radius at LO (see main text). 
The uncertainties are one-standard-deviation estimates which include the statistical Monte Carlo error as well as the errors due 
to the $N_t^{} \to \infty$ extrapolation.
\label{tab_trans}}
\end{center}
\vspace{-5mm}
\end{table}
With the scaling factor included, we find that the NLEFT predictions for the $E2$ and 
$E0$ transitions are in good agreement with the experimental values. 
In particular, NLEFT is able to explain the empirical value of $B(E2,2^+_1 \to 0^+_2)$, which is $\simeq 30$ times larger than the Weisskopf 
single-particle shell model estimate. This provides further confirmation of the interpretation of the $2^+_1$ state as a rotational excitation of the 
$0^+_2$ state. Finally, we provide a prediction for the quadrupole moment of the $2^+_1$ state. 

\begin{figure}[t]
\begin{center}
\includegraphics[width=.24\columnwidth]{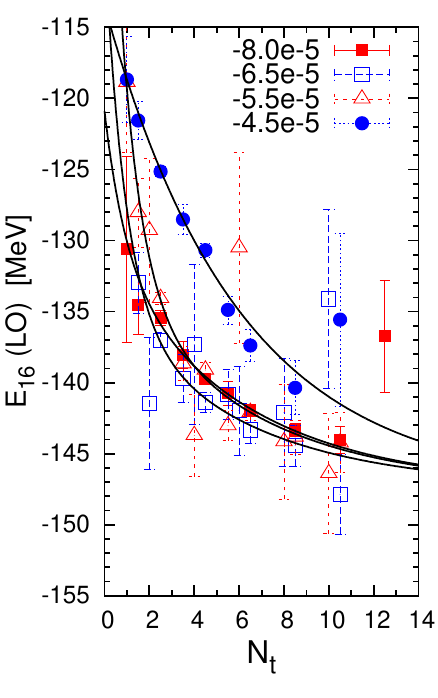}
\includegraphics[width=.24\columnwidth]{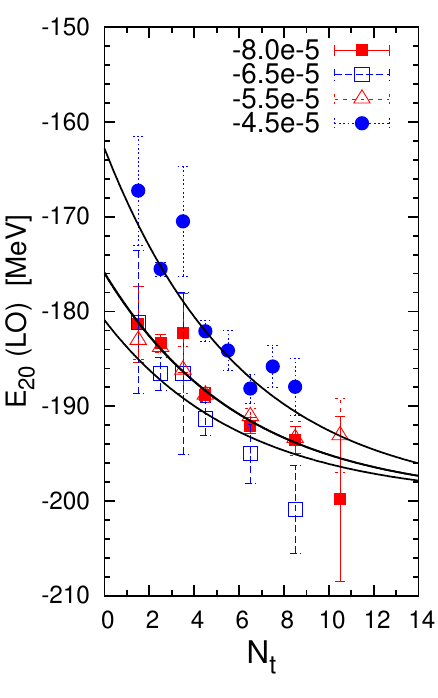}
\includegraphics[width=.24\columnwidth]{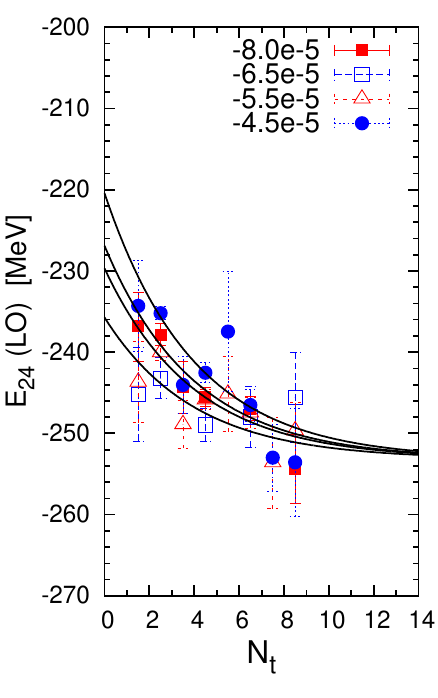}
\includegraphics[width=.24\columnwidth]{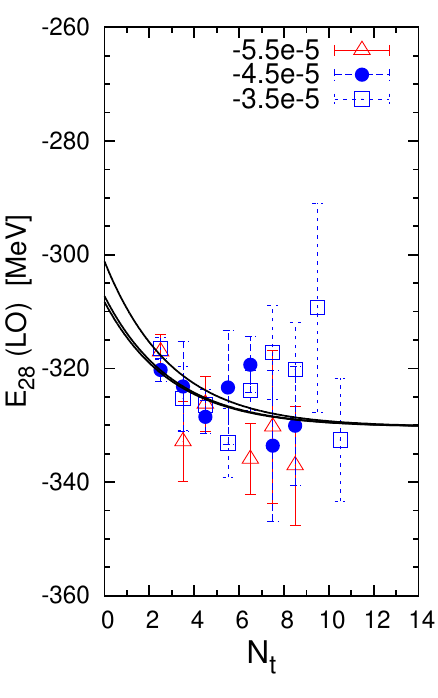}
\end{center}
\vspace{-6mm}
\caption{NLEFT results for the LO transient energy $E_A^{}(t)$ for $A = 16, 20, 24, 28$, with $C_\mathrm{SU(4)}^{}$ given (in MeV$^{-2}$) for each trial state, cf. 
Eq.~(\protect\ref{eq:H_SU4}).  For definitions, see  Ref.~\cite{Lahde:2013uqa}.
\label{fig:LO}}
\end{figure}

In Ref.~\cite{Lahde:2013uqa}, we addressed the questions: How large a nucleus can be calculated from first principles using the framework of 
chiral nuclear EFT, and what are the remaining challenges?   For finding answers to these questions,  we calculated the ground states of $\alpha$-type 
nuclei from $^{4}$He to $^{28}$Si.  We note that  NLEFT differs from other \textit{ab initio} methods in that it is an unconstrained 
Monte Carlo calculation, which does not require truncated basis expansions, many-body perturbation theory, or any constraint on the nuclear wave 
function. Thus, our NLEFT calculations are truly unbiased Monte Carlo simulations, and the results obtained and presented here are an important benchmark for 
\textit{ab initio} calculations of larger nuclei using chiral nuclear EFT. Any deficiencies are indicative of shortcomings in the specific nuclear 
interactions, rather than of errors generated by the computational method. Such a definitive analysis would be difficult to achieve
using other methods.   In Ref.~\cite{Lahde:2013uqa}, a new method to better perform the extrapolation to large Euclidean time in NLEFT
was suggested, denoted as ``triangulation''. The idea is to start from a set of  inexpensive low-energy SU(4) symmetric filters that differ by 
the strength of the SU(4) coupling, given by 
\begin{equation}
H_\mathrm{SU(4)}^{} \equiv H_\mathrm{free}^{} 
+\frac{1}{2} C_\mathrm{SU(4)}^{}\sum_{\vec n,\vec n'} 
{:\rho(\vec n)f(\vec n - \vec n')}\rho(\vec{n}'):~,
\label{eq:H_SU4}
\end{equation}
where $f$ is a Gaussian smearing function with its width set by the average effective range of the two S-wave interaction channels,
$\vec{n}, \vec{n}'$ are  points on the lattice and $\rho$ is  the total nucleon density.  For a precise definition of this SU(4) filter and its
role in performing nuclear lattice simulations, see~\cite{Borasoy:2006qn}. In Fig.~\ref{fig:LO}, the LO ground state energies for 
$^{16}$O, $^{20}$Ne,  $^{24}$Mg and  $^{28}$Si are shown for three or four  trial SU(4) symmetric states with different (unphysical)
coupling constants $C_{\rm SU(4)}$.
One clearly observes that 
this methods works and it is furthermore encouraging to note that these new extrapolations are consistent with our earlier results for $^{12}$C in 
Refs.~\cite{Epelbaum:2012qn,Epelbaum:2012iu}, which were computed using delocalized plane-wave as well as 
$\alpha$-cluster trial wave functions.
As can be seen from Fig.~\ref{fig:LO}, there is some overbinding at LO that increases with increasing atomic number $A$. This
feature persists up to NNLO, as shown in Tab.~\ref{tab:alpha}. This clearly is a systematic effect that can be traced back to some
deficiencies of our soft NNLO forces. In this context, we note that other \textit{ab initio} methods using 
soft potentials encounter similar problems in the description of both light and medium-mass nuclei using the same set of 
interactions~\cite{Hagen:2012fb,Jurgenson:2013yya,Roth:2011ar}.  For making progress, it is useful to explore 
the nature of the missing physics. As we ascend the $\alpha$ ladder from $^{4}$He to $^{28}$Si, the lighter nuclei can be described as collections of 
alpha clusters~\cite{Epelbaum:2012qn,Epelbaum:2011md}. As the number of clusters increases, they become increasingly densely packed, such that
a more uniform liquid of nucleons is approached. This increase in the density of $\alpha$-clusters appears correlated with the gradual overbinding 
we observe at NNLO for $A \geq 16$. As this effect becomes noticeable for $^{16}$O, we can view it as a problem which first
arises in a system of four $\alpha$-clusters. Following Refs.~\cite{Epelbaum:2009pd,Epelbaum:2010xt}, which removed discretization errors associated with four 
nucleons occupying the same lattice site, we can attempt to remove similar errors associated with four $\alpha$-clusters in close proximity on
 neighboring lattice sites. 
\begin{table}[tb]
\begin{center}
\begin{tabular}{|l|ccccccc|}
\hline
$A$                                           &      4             &      8      &      12      &      16      &      20     &      24      &      28       \\
\hline
E(LO)                                     &  28.87(6)    &   57.9(1)  &  96.9(2)  &  147.3(5) & 199.7(9) & 253(2) & 330(3)  \\  
E(NNLO)                               &  28.93(7)    &   56.4(2)  &  91.7(2)  &  138.8(5) &  184.3(9) & 232(2) & 308(3)   \\
E(NNLO+4N$_{\rm eff})$   &  28.93(7)    &   56.3(2)  &  90.3(2)  &  131.3(5) &  165.9(9) & 198(2) & 233(3)  \\
\hline
Exp.                                          &   28.30        &  56.35   &   91.26   &   127.62   &    160.64   & 198.26 & 236.54  \\
\hline
\end{tabular}
\caption{
NLEFT results for the ground-state energies  of the alpha-cluster nuclei. 
The combined statistical and extrapolation errors are given in parentheses. The rows labelled E(LO) and E(NNLO)
give the results at LO and NNLO, in order. 
The row E(NNLO+4N$_{\rm eff}$) includes the effective 4N contribution from 
Eq.~(\protect\ref{eq:4N}). 
Finally, the row Exp.  gives the empirical energies.  All energies are in MeV.
\label{tab:alpha}}
\end{center}
\end{table}
The simplest interaction which permits a removal of the overbinding associated with such configurations is 
of the form
\begin{equation}
V^{(4\mathrm{N_{eff}})}
 = D^{(4\mathrm{N_{eff}})} \hspace{-.6cm}
\sum_{1\le(\vec n_{i}^{}-\vec{n}_{j}^{})^2_{}\le2} \hspace{-.5cm}
\rho(\vec n_1^{})\rho(\vec n_2^{})\rho(\vec n_3^{})\rho(\vec n_4^{}), \label{eq:4N}
\end{equation}
with $\rho(\vec{n})$ the total nucleon density. The summation includes nearest or 
next-to-nearest neighbor (spatial) lattice sites. One can tune $D^{(4\mathrm{N_{eff}})}$  e.g. to give approximately the correct energy for the ground state of $^{24}$Mg. With  $V^{(4\mathrm{N_{eff}})}$ included, a good description of the ground-state energies is obtained over the full range from light 
to medium-mass nuclei, with a maximum deviation from experiment no larger than $\sim 3$\%, see Tab.~~\ref{tab:alpha}.
This lends support to the qualitative picture that the overbinding of the NNLO results in 
Table~\ref{tab:alpha} is associated with the increased packing of $\alpha$-clusters and the eventual crossover to a uniform nucleon liquid.  
The missing physics would then be comprised of short-range repulsive forces that counteract the dense packing of alpha clusters. 
From this analysis, the path forward for {\it ab initio} calculations of heavier nuclei using chiral nuclear EFT appears clear. 
The softening of the two-nucleon interaction should not be pushed so far that heavier nuclei become significantly over-bound by the 
two-nucleon force alone. This is not merely an issue for NLEFT, but would appear to be a universal criterion for all {\it ab initio} methods.  
A concerted effort should be made to improve the current computational algorithms to handle interactions with more short-range repulsion. 
The NLEFT collaboration is now exploring this approach for simulations of larger nuclei. We are now in the process of improving the 
lattice algorithms for simulations at smaller lattice spacings, and extending NLEFT to N3LO in the chiral expansion.

\begin{figure}[t]
\includegraphics[width=\textwidth]{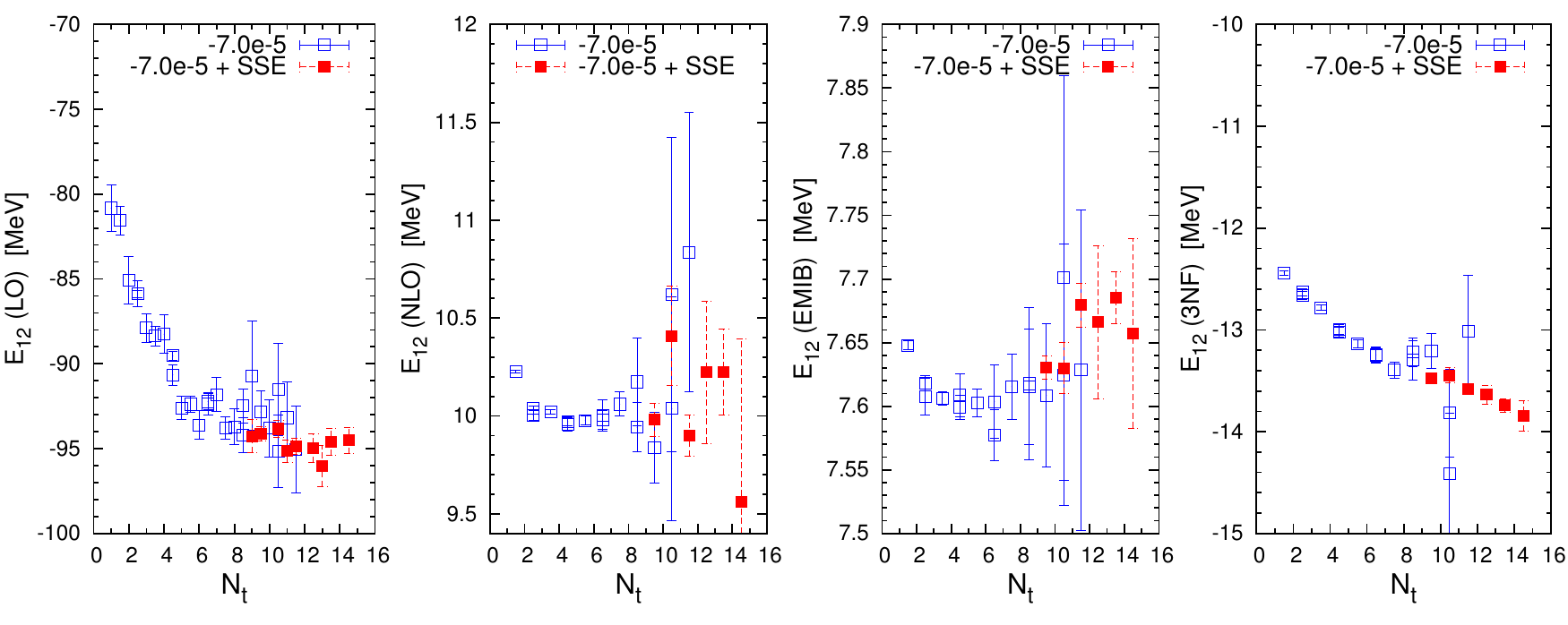}
\caption{\label{fig:12C_v11} Comparison of the new PMC data for $^{12}$C from the SSE analysis (red filled squares) and 
previous calculations~\cite{Lahde:2013uqa}  for $d_h^{} = 1$ (blue open squares). 
The contributions from the improved leading order 
amplitude~(LO), the two-nucleon force at next-to-leading order~(NLO), the electromagnetic and strong isospin breaking~(EMIB) and the three-nucleon force at next-to-next-to-leading order~(3NF) are shown separately. The results correspond to a trial state with an SU(4) coupling of $-7.0 \times 10^{-5}$~MeV$^{-2}$, 
not to be confused with the SU(4) coupling $C_{\rm SU(4)}^{}$ for the SSE analysis. It should be noted that the exponential deterioration of the Monte Carlo error has been circumvented. Note these data should not be interpreted in terms of a plateau. For more details, see Ref.~\cite{Lahde:2015ona}.}
\end{figure}

Projection Monte Carlo (PMC) calculations of lattice chiral EFT suffer from sign oscillations to a varying degree dependent on
the number of protons and neutrons, which are minimal for the $\alpha$-type nuclei discussed so far. 
In Ref.~\cite{Lahde:2015ona}, we have introduced the ``symmetry-sign extrapolation'' (SSE) method, 
which allows us to use the approximate Wigner SU(4) symmetry of the nuclear interaction to systematically 
extend the  PMC calculations to nuclear systems where the sign problem is severe.
For that, one defines the ``interpolating Hamiltonian'' $H$ as
\begin{equation}
H \equiv d_h^{} H_\mathrm{LO}^{} + (1-d_h^{}) H_{\rm SU(4)},
\label{Hd}
\end{equation}
which depends on $d_h^{}$ as well as the (unphysical) coupling constant $C_{\rm SU(4)}^{}$, cf.
Eq.~(\ref{eq:H_SU4}). 
This can also be viewed as giving the interaction parameters a linear dependence on $d_h^{}$.
By taking $d_h^{} < 1$, we can always decrease the sign problem to a tolerable level, while simultaneously tuning 
$C_{\rm SU(4)}^{}$ to a value favorable for an extrapolation $d_h^{} \to 1$. Most 
significantly, we can make use of the constraint that the physical result at $d_h^{} = 1$ should be independent of $C_{\rm SU(4)}^{}$. 
The dependence of calculated matrix elements on $d_h^{}$ is smooth in the vicinity of $d_h^{} = 1$. 
We note that an extrapolation technique similar to SSE has been used in Shell Model Monte Carlo calculations for over two decades~\cite{Alhassid:1993yd,Koonin:1996xj}.  In that case, the extrapolation is performed by decomposing the Hamiltonian into ``good sign'' and ``bad sign'' parts, 
$H_G^{}$ and $H_B^{}$, respectively. The calculations are then performed 
by multiplying the coefficients of $H_B^{}$ by a parameter $g$ and extrapolating from $g < 0$, where the simulations are free from sign 
oscillations, to the physical point $g = 1$. For SSE, the analysis in terms of  ``good'' and ``bad'' signs is not the entire story. 
Most of the interactions can be divided into two groups which are ``sign free'' by themselves, such that a large 
portion of the sign oscillations is due to interference between the different underlying symmetries of the two groups of interactions. 
Since this effect is quadratic in the interfering interaction  coefficients, the growth of the sign problem is more gradual. We therefore expect to be able to extrapolate from values not so far away from the physical point $d_h^{} = 1$.
We  first discuss results for the $^{12}$C nucleus, as this provides us with a convenient test case for the SSE method.
We extend the PMC calculation to larger values of the Euclidean projection time than would otherwise be possible, and 
verify how well these new results agree with earlier calculations. Thus, we shall work at finite Euclidean time, extrapolate such data to 
$d_h^{} \to 1$, after which the extrapolation $L_t^{} \to \infty$ is performed. For $^{12}$C, we have performed PMC calculations using SSE for Euclidean projection times between $N_t^{} = 9.0$ and~$14.5$ for one of the trial states
used in Ref.~\cite{Lahde:2013uqa}. It should be noted that the results of Ref.~\cite{Lahde:2013uqa} correspond to $d_h^{} = 1$, and could thus only be extended to $N_t^{} \simeq 10$ before the
sign problem became prohibitive. In Fig.~\ref{fig:12C_v11}, we extend the dataset of Ref.~\cite{Lahde:2013uqa} with the new SSE data.
This clearly  establishes that the SSE data are consistent with those obtained at $d_h^{} = 1$, when such calculations are not
prohibited by the sign problem,  for more details, see Ref.~\cite{Lahde:2015ona}.

The sign problem in the $A = 6$ system with 2~protons and 4~neutrons (or {\it vice versa}) is somewhat more severe than for $^{12}$C. Hence, if calculations are performed entirely at $d_h^{} = 1$, the extrapolation to infinite Euclidean time (while still feasible) has to be performed using data with a rather limited range in $N_t^{}$.  However, for $d_h^{} < 1$ this situation improves rapidly. For $^6$He and $^6$Be, one should therefore approach the problem differently than for $^{12}$C. We perform the extrapolation in Euclidean time for each pair of $C_{\rm SU(4)}^{}$ and $d_h^{}$, with the extrapolation $d_h^{} \to 1$ as the final step of the analysis.  A detailed discussion of the simulations performed in these systems is given in Ref.~\cite{Lahde:2015ona}. Here, we just add some remarks.
The SSE method introduced  is inspired by the existence of an SU(4) symmetric Hamiltonian which provides a reasonably accurate description of the
physics of the full NLEFT Hamiltonian. This has already proven useful in the work discussed above, as it greatly facilitates finding an accurate initial wave function which minimizes the extent of Euclidean time projection necessary with the full Hamiltonian. With the SSE,  this concept can be taken one step further, by studying a weighted sum of physical and SU(4) symmetric Hamiltonians. In this way, the sign problem could be arbitrarily ameliorated, at the price of introducing an extrapolation in a control parameter $d_h^{} \to 1$. In practice, this means that the SSE method is only useful as long as the extrapolation errors can be kept under control. Naturally, performing simulations at a range of values of $d_h^{}$ has the potential to multiply the required CPU time by a large factor. However, we have found that we are able to avoid an exponential increase in computation time as a function
of Euclidean projection time, as long as we are able to perform simulations for $d_h^{} > 0.75$, as the accuracy of extrapolation then remains comparable with the statistical errors of  typical simulations at $d_h^{} = 1$. We have also explored the freedom in the choice of the SU(4) symmetric Hamiltonian, which clearly plays no role at $d_h^{} = 1$, but which in general gives different results for $d_h^{} \neq 1$. We have therefore made use of a ``triangulation'' method 
to improve the accuracy of the extrapolation $d_h^{} \to 1$. An important consideration is whether a continuous shift from an SU(4) symmetric Hamiltonian 
to the full chiral EFT Hamiltonian can be effected without inducing 
non-trivial changes in the spectrum. For instance, the appearance of a level crossing at a critical value of $d_h^{}$
would clearly limit the applicability of SSE. We note that such level crossings as a function of $d_h^{}$ would be quite rare for low-lying 
nuclear bound states, and furthermore we  have the freedom to choose $C_{\rm SU(4)}$ to avoid such level crossings. But if a level crossing were 
to occur, there would be some subtleties in obtaining accurate and converged results.   For such cases, it would be preferable to take the 
$d_h^{} \to 1$ limit first, followed by extrapolation in Euclidean time. An even better solution would entail solving a coupled-channel problem
using multiple initial states. This makes it possible to  disentangle one or more nearly degenerate states with the same quantum numbers. 
So far, we have concentrated on systems where such degeneracies are  not expected, and where PMC is still possible (though difficult) 
without the SSE method. One also should explore different ``extrapolation Hamiltonians'', where the sign oscillations are 
minimized while retaining as much as possible of the full chiral EFT structure.

\section{Further developments}

Here, I summarize some further important developments which are partly covered by the talks in this session, and
provide additional references:
\begin{itemize}
\item{\bf Lattice spacing dependence:}~Most of the simulations reported here have been performed on a coarse lattice with a lattice spacing 
$a = 1.97\,$fm. This raises the question whether there are sizeable lattice artifacts related to this spacing. An important first step in this direction
is the  study on the lattice spacing dependence in Ref.~\cite{Klein:2015vna}.  
There, the two-body system for lattice spacings
from $a=0.5\,$fm to $a=2.0\,$fm at lowest order in the pionless as well as in the pionful theory
was investigated. In the pionless case, a simple Gaussian smearing allows to demonstrate lattice spacing independence over a wide range of lattice spacings. It was also shown that regularization methods 
known from the continuum formulation \cite{Epelbaum:2014efa} are necessary as well as feasible for the pionful approach and will lead to 
$a$-independent results. In $\alpha$-cluster models, the $a$-dependence of
various observables was also studied, see Refs.~\cite{Lu:2014xfa,Lu:2015gfa}.
For further work on this issue, see the contribution from Alarcon to these proceedings~\cite{JMA}.
\item{\bf Breaking and restoration of rotational symmetry:}~On a periodic
  lattice, the rotational symmetry SO($3$) is broken down to the cubic group SO($3,Z$),
  that has five irreducible representations. In Ref.~\cite{Lu:2014xfa} the
  breaking of rotational symmetry on the lattice for bound state energies and
  practical methods for suppressing this breaking were explored. The general
  problems associated with lattice discretization errors and finite-volume
  errors were discussed using an $\alpha$-cluster model for $^8$Be and
  $^{12}$C.  The focus was put on the lowest states with non-zero angular
  momentum which split into multiplets corresponding to different irreducible
  representations of the cubic group. The dependence of such splittings on the
  lattice spacing and the box size was examined. The lattice spacing errors
  are closely related to the commensurability of the lattice with the
  intrinsic length scales of the system.  Rotational symmetry breaking effects 
  can be significantly reduced by using improved lattice actions, and  the
  physical energy levels are accurately reproduced by the weighted average of
  a given spin multiplets. This was extended to the study of irreducible
  tensor operators in Ref.~\cite{Lu:2015gfa}. The lowest states with non-zero
  angular momentum  were considered and the matrix elements of multipole
  moment operators were examined. The physical reduced matrix element is well 
  reproduced by averaging over all possible orientations of the quantum state, 
  and this is expressed as a sum of matrix elements weighted by the
  corresponding Clebsch-Gordan coefficients. For our $\alpha$-cluster model, 
  we find that the effects of rotational symmetry breaking can be largely
  eliminated for lattice spacings of $a\leq 1.7\,$fm, and we expect similar 
  improvement for actual lattice Monte Carlo calculations. For related work 
  on the restoration of rotational symmetry in the continuum limit of lattice 
  field theories see~\cite{Davoudi:2012ya}.
\item{\bf Discretized forces to higher orders and extraction of NN phase
    shifts:}~Clearly, the forces at NNLO used so far are not sufficiently accurate,
    as witnessed e.g. by the overbinding problem discussed before. This is to some
    extent also related to the extraction of the nucleon-nucleon phase shifts 
    based on the so-called spherical wall method, that was developed for NLEFT
    in Ref.~\cite{Borasoy:2007vy} (see also Ref.~\cite{Carlson:1984zz} 
    for a similar approach). 
    This method is better suited for nuclear physics as it  deals much better 
    with partial-wave mixing than L\"uscher's well-known approach. However, to
    further improve the precision of the  determination of phase shifts and mixing 
    angles on the lattice, a novel approach was presented in Ref.~\cite{Lu:2015riz}.
    It starts with angular momentum projection on the lattice wave functions in 
    order to construct lattice radial wave functions. One then imposes spherical wall 
    boundaries together with an adjustable auxiliary potential to determine 
    phase shifts at arbitrary energy,
\begin{equation}
        V_{{\rm aux}}(r) \equiv
       V_0 \exp\left(-(r-R_{W})^{2}/a^{2}\right),\quad R_{0}\leq r\leq R_{W},
\label{eq:auxiliarypotential}
\end{equation}
    where $R_W$ is the wall boundary, $V_0$ the strength of the auxiliary potential,
    $a$ is the lattice spacing and $R_0$ is chosen to be larger than the range
    of the interaction but smaller than $R_W$. For coupled partial waves, a 
    complex-valued auxiliary potential is used that breaks time-reversal invariance 
    in order to extract phase shifts and mixing angles from the complex-valued wave 
    functions. The method was benchmarked using a system of two spin-1/2 particles 
    interacting through a finite-range potential with a strong tensor component, also
    used in Ref.~\cite{Borasoy:2007vy}. 
    Scattering parameters can be extracted precisely for all angular momenta and 
    energies, see Fig.~\ref{fig:wall}. Using this new method and an improved smearing,
    the NLEFT collaboration is presently working out the forces to N3LO.
\begin{figure}[h]
  \begin{center}
\includegraphics[width=0.95\textwidth]{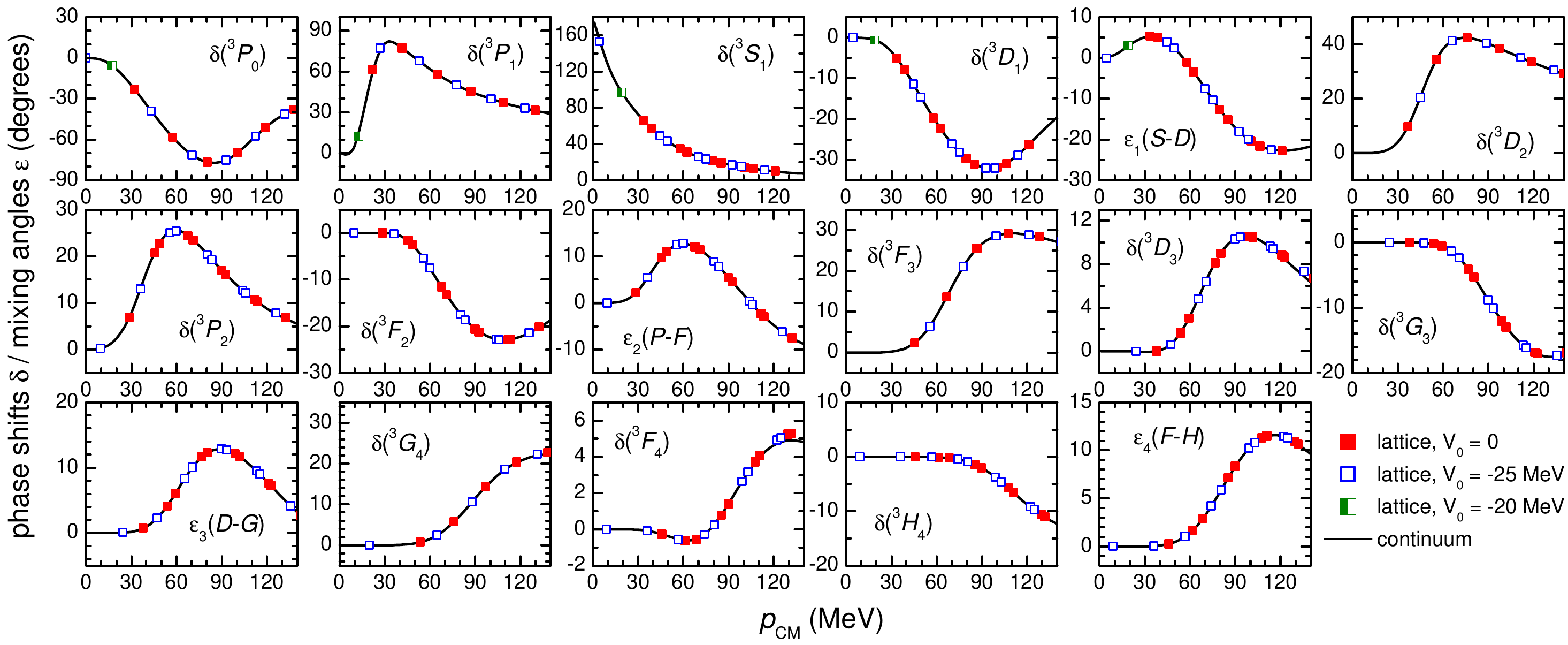}\hspace{1pc}
\end{center}
\vspace{-2mm}
\caption{\label{fig:wall}
 Phase shifts and mixing angles for $J\leq4$ in the $S=1$ channel. Full, open 
and half-open squares denote the results obtained with auxiliary potential 
strength $V_0=0$, $V_0=-25$~MeV and $V_0=-20$~MeV, respectively.
For $V_0=-20$ MeV only selected results are shown.
Solid lines denote continuum results. For details, see Ref.~\protect\cite{Lu:2015riz}.}
\vspace{-2mm}
\end{figure} 

\item{\bf Ab initio calculation of scattering and inelastic processes:}~Scattering
  processes and inelastic reactions like radiative neutron capture on the proton are
  of major importance to understand the generation of the element in the Big Bang
  and in stars.   Such processes can be tackled using the recently
  proposed adiabatic projection method (APM) that allows to construct an effective
  cluster Hamiltonian from the underlying chiral EFT Hamiltonian, 
  see Refs.~\cite{Rupak:2013aue,Pine:2013zja} and the talk by Rokash~\cite{AR}. This method 
  is similar to the NCSM/RGM approach to nuclear scattering developed in
  Ref.~\cite{Quaglioni:2009mn}.   We are presently working on various scattering reactions.
\end{itemize}

\section{Outlook}

NLEFT has enjoyed a number of remarkable successes like the Hoyle state calculation,
the investigations of the fine-tunings in the triple-alpha process or the {\em ab initio}
calculation of $^{16}$O based on the chiral EFT forces at NNLO.
After these promising results, it is now time to consolidate this framework in the sense
that the remaining inaccuracies of the underlying forces should be eliminated and  more 
detailed investigations of nuclear structure and dynamics have to be performed. Just 
to mention the rich experimental information from electron scattering, the quest to
understand the limits of the nuclear stability or the generation of elements in the
Big Bang and in stars through nuclear reactions. Time is ripe to make real progress in 
all these areas. This requires more manpower than provided by the NLEFT collaboration 
and I hope that the physics presented here will stimulate other groups to use and further
develop this powerful tool.

\acknowledgments

I thank my collaborators Dean Lee, Timo L\"ahde, Tom Luu, Bingnan Lu, Ning Li, 
Serdar Elhatisari, Nico Klein, Evgeny Epelbaum, Hermann Krebs, Gautam Rupak 
and Jos\'e Alarcon for sharing their insights into the topics discussed here.



\end{document}